\documentclass[twocolumn, secnumarabic,amssymb, amsmath, nofootinbib, aps, prd, twoside]{revtex4-2}
\usepackage{xurl}
\usepackage{natbib}
\usepackage[utf8]{inputenc}
\usepackage{fancyhdr}
\usepackage{academicons}
\usepackage[dvipdfm]{graphicx}
\usepackage{xcolor}
\usepackage{subcaption} 
\usepackage{hyperref}

\usepackage{tikz}
\usepackage{slashed}
\usepackage{tikz-3dplot}
\tdplotsetmaincoords{70}{50}
\usepackage{graphicx}
\usepackage{subcaption} 

\pgfmathsetmacro{\rvec}{2.5}
\pgfmathsetmacro{\thetavec}{45}
\pgfmathsetmacro{\phivec}{60}
\usepackage{xcolor}
\definecolor{dodgerblue}{rgb}{0.12, 0.56, 1.0}
\usetikzlibrary{positioning,arrows}
\usetikzlibrary{decorations.pathmorphing}
\usetikzlibrary{decorations.markings}
\tikzset{
	particle/.style={thick,draw=black, postaction={decorate},
		decoration={markings,mark=at position .5 with {\arrow{latex}}}},
	gluon/.style={decorate, draw=black,
		decoration={coil,aspect=0,segment length=5.8pt,amplitude=5pt}}
}
\usetikzlibrary{positioning,arrows}
\usetikzlibrary{decorations.pathmorphing}
\usetikzlibrary{decorations.markings}
\usetikzlibrary{patterns,snakes}
\tikzset{
	particle/.style={thick,draw=black, postaction={decorate},
		decoration={markings,mark=at position .5 with {\arrow{latex}}}},
	gluon/.style={decorate, thick, draw=orange,
		decoration={coil,aspect=0,segment length=5.8pt,amplitude=5pt}}
}

\begin{document}


\title{Euclidean Methods on Dirac Quantization and Gauge Representation}


\author{Takashi Kiryu}
\email[]{u685087j@ecs.osaka-u.ac.jp}
\affiliation{Department of Physics, Osaka University, Osaka 560-0043, Japan}


\date{\today}

\begin{abstract}
Geometrical interpretation on U(1) gage theory of Dirac monopole, introduced here from the line integral\cite{Brandt} form of vector potentials, shows the gauge representation be multi-valued. In this paper, we construct Euclidean form of U(1) gauge theory on Dirac monopole based on the geometrical interpretation, derive the Dirac quantization condition geometrically. Even several theoretical methods which leads to Dirac quantization $eg=l/2$ under the presence of single magnetic pole has been developed, these are only possible under the postulate that gauge representation is single-valued. Using the line integration representation of vector potential, we derived the gauge equation which satisfies under the `phase gauge transformation' ($\mathbf{A}\to\mathbf{A}'=\mathbf{A}, \ \psi\to\psi'=e^{ie\Lambda}\psi$), also showed that the Dirac quantization condition can be derived by the geometrical boundary condition on this gauge equation, without any supposing of canonical quantization condition.  We finally explain the geometrical meaning of constant phase term of gauge and why the gauge be single-valued.
\end{abstract}


\maketitle

\section{introduction}
On the Dirac's paper\cite{taro}, the vector potential of single magnetic pole of the $+z$-axis string $\mathbf{n}(0,0,1)$and $-z$-axis string $\mathbf{n}'$(0,0,-1) is given by
\begin{equation}
\begin{aligned}
\mathbf{A}_{\mathbf{n}}&=-g\frac{1+\cos{\theta}}{r\sin{\theta}}\ \mathbf{e}_\varphi,\\[2ex]
\mathbf{A}_{\mathbf{n}'}&=g\frac{1-\cos{\theta}}{r\sin{\theta}}\ \mathbf{e}_\varphi.
\end{aligned}
\end{equation}
This gives representation of magnetic fields each potential $\mathbf{A}_{\mathbf{n}}(\mathbf{r})$ and $\mathbf{A}_{\mathbf{n}'}(\mathbf{r})$ makes using the regularized from of potential\cite{monopole, DD, A,  Lee} for the exclusion of singularity as
\begin{equation}
\begin{aligned}
\mathbf{B}_{\mathbf{n}}(\mathbf{r})=g\frac{\mathbf{r}}{r^3}-4g\pi\mathbf{n}\theta(z)\delta(x)\delta(y),\\[2ex]
\mathbf{B}_{\mathbf{n}'}(\mathbf{r})=g\frac{\mathbf{r}}{r^3}-4g\pi\mathbf{n}'\theta(z)\delta(x)\delta(y)
\end{aligned}\label{2}
\end{equation}
where we put $\theta(z)$ as Heavside step function. From the representation of magnetic field \eqref{2}, Dirac string($\mathbf{n}$ or $\mathbf{n}'$) has both divergent field $\mathbf{B}_g(\mathbf{r})$ and string field $\mathbf{B}_{\rm str}(\mathbf{r})$. Hereby the string field only holds the Bianchi identity of vector potential as $\boldsymbol{\nabla}\cdot\mathbf{B}=4g\pi-4g\pi=0$. This gives the line integral representation\cite{monopole, DD, A} along to the string from the Green function of poisson equation as
\begin{equation}\label{3}
\begin{aligned}
\mathbf{A}_{\mathbf{n}}(\mathbf{r})&=\int_{\tau\to\infty}\boldsymbol{\nabla}\left(\frac{1}{|\mathbf{r}-\mathbf{r}'|}\right)\times\mathbf{B}_{\rm str}\ d^3\mathbf{r}'\\[2ex]
&=g\int_{C(\mathbf{n})} \frac{(\mathbf{r}-\mathbf{r}')\times d\mathbf{r}'}{|\mathbf{r}-\mathbf{r}'|^3}
\end{aligned}
\end{equation}
where we put the integral path $C(\mathbf{n})$ as string path which starts from the single magnetic pole $g$. Therefore the U(1) gauge transformation gives rotation of string $\mathbf{n}\to\mathbf{n}''$ also represented by line integration on same integral path $C(\mathbf{n})$. Hereby we put the solid angle $\Omega_{\mathbf{n}\to\mathbf{n}''}(\mathbf{r})$ on gauge transformation $\mathbf{A}_{\mathbf{n}}(\mathbf{r})\to\mathbf{A}_{\mathbf{n}''}(\mathbf{r})$ as:
\begin{figure}[!h]
\centering
\includegraphics[width=\linewidth]{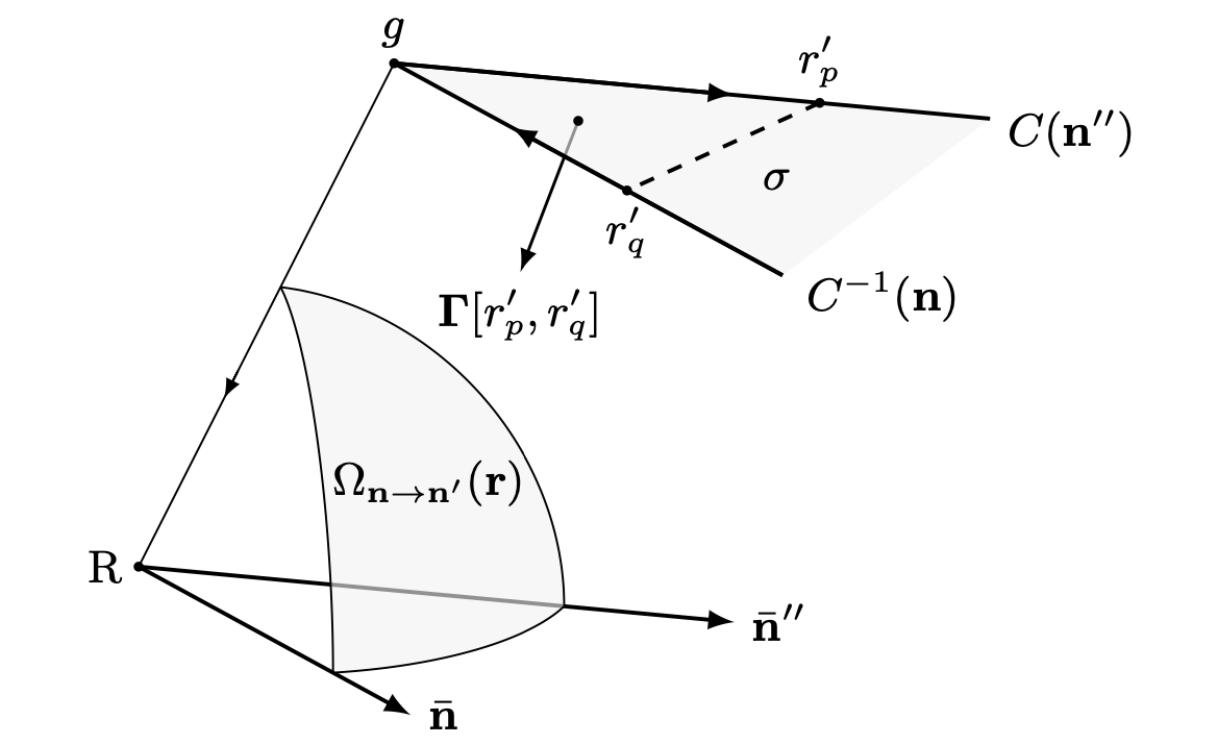}
\caption{The solid angle on vector $\mathbf{r}$ of string($\mathbf{n}$ and $\mathbf{n}''$).}
\label{1a}
\end{figure}\\
to relating geometrical rotation of string and U(1) gauge transformation. Here the direction of string-transforming area vector $\boldsymbol{\Gamma}[r_p', r_q']$ is define as which follows the Stoke's theorem on closed path $g\to r_p'\to r_q'$. Then the solid angle $\Omega_{\mathbf{n}\to\mathbf{n}''}(\mathbf{r})$ on {FIG. \ref{1a}} is represented by
\begin{equation}\label{4}
\Omega_{\mathbf{n}\to\mathbf{n}''}(\mathbf{r})=\int_{\sigma} \frac{(\mathbf{r}-\mathbf{r}')\cdot d\boldsymbol{\sigma}'}{|\mathbf{r}-\mathbf{r}'|^3}.
\end{equation}
Now we come to the U(1) gauge representation of string rotation $\mathbf{n}\to\mathbf{n}''$. Let us put the U(1) gauge transformation of fields $\psi_{\mathbf{n}}(\mathbf{r})$ and $\mathbf{A}_{\mathbf{n}}(\mathbf{r})$ on string $\mathbf{n}$ for the string rotation $\mathbf{n}\to\mathbf{n}''$ as:
\begin{equation}
\begin{aligned}
\psi_{\mathbf{n}}(\mathbf{r})\ &\to\ \psi_{\mathbf{n}''}(\mathbf{r})=e^{ie\Lambda(\mathbf{r})}\psi_{\mathbf{n}}(\mathbf{r}),\\[2ex]
\mathbf{A}_{\mathbf{n}}(\mathbf{r})\ &\to\ \mathbf{A}_{\mathbf{n}''}(\mathbf{r})=\mathbf{A}_{\mathbf{n}}(\mathbf{r})+\boldsymbol{\nabla}\Lambda(\mathbf{r})
\end{aligned}
\end{equation}
where the notation $\psi_{\mathbf{n}''}(\mathbf{r})$ and $\mathbf{A}_{\mathbf{n}''}(\mathbf{r})$ show the fields on rotated string $\mathbf{n}''$. If we observe the line integration representation of vector potential of each string $\mathbf{n}, \ \mathbf{n}''$ on \eqref{3}, the gauge transformation of vector potential now comes to
\begin{equation}\label{6}
\delta\mathbf{A}_{\mathbf{n}}(\mathbf{r})=g\int_{C(\mathbf{n}'')+C^{-1}(\mathbf{n})}\frac{(\mathbf{r}-\mathbf{r}')\times d\mathbf{r}'}{|\mathbf{r}-\mathbf{r}'|^3}=\boldsymbol{\nabla}\Lambda(\mathbf{r})
\end{equation} 
where $\delta\mathbf{A}_{\mathbf{n}}(\mathbf{r})=\mathbf{A}_{\mathbf{n}''}(\mathbf{r})-\mathbf{A}_{\mathbf{n}}(\mathbf{r})$. As boundary integral on $[r_p', r_q']$ vanished, we shall use the modified closed path of string $\boldsymbol{\Gamma}[r_p', r_q']$ on integration \eqref{6}, the Stoke's theorem gives directly:
\begin{equation}\label{7}
\begin{aligned}
\delta\mathbf{A}_{\mathbf{n}}(\mathbf{r})&=\lim_{r_p', r_q'\to\infty}g\int_{\partial\boldsymbol{\Gamma}[r_p', r_q']}\frac{(\mathbf{r}-\mathbf{r}')\times d\mathbf{r}'}{|\mathbf{r}-\mathbf{r}'|^3}\\[2ex]
&=\lim_{r_p', r_q'\to\infty}-g\boldsymbol{\nabla}\times\int_{\boldsymbol{\Gamma}[r_p', r_q']}\boldsymbol{\nabla}\left(\frac{1}{|\mathbf{r}-\mathbf{r}'|}\right)\times d\boldsymbol{\sigma}'\\[2ex]
&=g\left[\boldsymbol{\nabla}\left(-\boldsymbol{\nabla}\cdot\int_{\sigma}\frac{d\boldsymbol{\sigma}'}{|\mathbf{r}-\mathbf{r}'|}\right)\right]\\[1ex]
& \ \ \ \ \ \ \ \ \ \ \ \ \ \ \ \ \ \ \ \ \ \ \ \ \ \ \ \ \ \ +g\int_{\sigma} \nabla^2\left(\frac{1}{|\mathbf{r}-\mathbf{r}'|}\right)\ d\boldsymbol{\sigma}'.
\end{aligned}
\end{equation}
On second line of calculation \eqref{7}, we use the Stoke's theorem in form of 
\begin{equation}\label{8}
\int_{\partial\Omega} f\ d\mathbf{r}=\int_{\Omega} \boldsymbol{\nabla}f\times d\boldsymbol{\sigma},
\end{equation}
also vector calculus identity $\boldsymbol{\nabla}\times(\boldsymbol{\nabla}\times\mathbf{A})=\boldsymbol{\nabla}(\boldsymbol{\nabla}\cdot\mathbf{A})-\nabla^2\mathbf{A}$ used on third line. By feeding \eqref{4} to \eqref{6} and \eqref{7}, we get generalized representation\cite{monopole, DD, A} of gauge $\Lambda(\mathbf{r})$ as
\begin{equation}\label{9}
\boldsymbol{\nabla}\Lambda(\mathbf{r})=g\boldsymbol{\nabla}\Omega_{\mathbf{n}\to\mathbf{n}''}(\mathbf{r})-4g\pi\int_\sigma d\boldsymbol{\sigma}'\ \delta^{(3)}(\mathbf{r}-\mathbf{r}').
\end{equation}
Now we can evaluate U(1) gauge $\Lambda(\mathbf{r})$. Since the gradient of solid angle $\boldsymbol{\nabla}\Omega_{\mathbf{n}\to\mathbf{n}''}(\mathbf{r})$ diverges on point $\mathbf{r}_\sigma\in\sigma$, we have observation of gauge equation \eqref{9} in form 
\begin{equation}\label{10}
\boldsymbol{\nabla}\left(\Omega_{\mathbf{n}\to\mathbf{n}''}(\mathbf{r})-\frac{1}{g}\Lambda(\mathbf{r})\right)=4\pi\int_\sigma d\boldsymbol{\sigma}'\ \delta^{(3)}(\mathbf{r}-\mathbf{r}').
\end{equation} 
Then, we shall consider two path $\gamma_1(t), \gamma_2(t)\ (0\leq t\leq 1)$ of line integration on \eqref{10} shown by:
\begin{figure}[!h]
\centering
\includegraphics[width=1\linewidth]{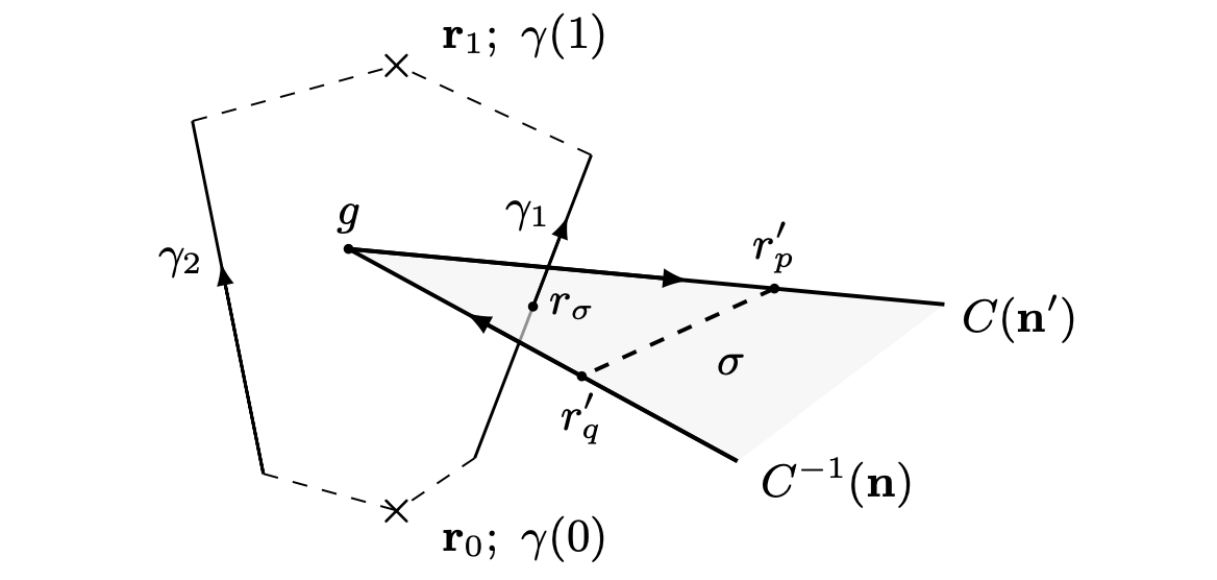}
\caption{The path $\gamma_1, \gamma_2$ of line integral on \eqref{10}}
\label{2a}
\end{figure}\\
which is homotope($\gamma_1(0)=\gamma_2(0), \gamma_1(1)=\gamma_2(1)$).
Each path of line integrals on \eqref{10} gives
\begin{equation}\label{11}
\begin{aligned}
\delta_1\left(\Omega_{\mathbf{n}\to\mathbf{n}''}(\mathbf{r})-\frac{1}{g}\Lambda(\mathbf{r})\right)=4\pi\iint_{\gamma_1\times \sigma} d\mathbf{r}\cdot d\boldsymbol{\sigma}'\ \delta^{(3)}(\mathbf{r}-\mathbf{r}'),\\[2ex]
\delta_2\left(\Omega_{\mathbf{n}\to\mathbf{n}''}(\mathbf{r})-\frac{1}{g}\Lambda(\mathbf{r})\right)=4\pi\iint_{\gamma_2\times \sigma} d\mathbf{r}\cdot d\boldsymbol{\sigma}'\ \delta^{(3)}(\mathbf{r}-\mathbf{r}').
\end{aligned}
\end{equation}
These results are supported by the integral term
\begin{equation}
\iint_{\gamma\times \sigma} d\mathbf{r}\cdot d\boldsymbol{\sigma}'\ \delta^{(3)}(\mathbf{r}-\mathbf{r}')
\end{equation}
converges.
Now we get this U(1) gauge $\Lambda(\mathbf{r})$ be discontinuous and multi-valued function on $\mathbf{r}\not\in\mathbf{r}_\sigma$. The $\gamma_1$ like path which passing through the string transforming area $\sigma$ gives integral term on right side of \eqref{11} as $-4\pi$. But, the $\gamma_2$ like path which passing through the outside of string transforming area $\sigma^c$ integral term on right side of \eqref{11} gives as 0. If we put gauge fixing on string rotation $\mathbf{n}\to\mathbf{n}''$ on FIG. \ref{2a} as $\psi(\mathbf{r}_0)\ \to\ \psi'(\mathbf{r}_0)=e^{ieg\Omega_{\mathbf{n}\to\mathbf{n}''}(\mathbf{r}_0)}\psi(\mathbf{r}_0)$, the gauge transformation on $\mathbf{r}_1$ have following two representations:
\begin{equation}\label{13}
\begin{aligned}
\psi(\mathbf{r}_1)\ \to\ \psi'(\mathbf{r}_1)&=U_1\psi(\mathbf{r}_1); \ \ \ U_1=e^{ieg(\Omega_{\mathbf{n}\to\mathbf{n}''}(\mathbf{r}_1)+4\pi)},\\[2ex]
\psi(\mathbf{r}_1)\ \to\ \psi'(\mathbf{r}_1)&=U_2\psi(\mathbf{r}_1); \ \ \ U_2=e^{ieg\Omega_{\mathbf{n}\to\mathbf{n}''}(\mathbf{r}_1)}.
\end{aligned}
\end{equation}
On the limit $\mathbf{r}_0\to\mathbf{r}_{\sigma}-$, $\mathbf{r}_1\to\mathbf{r}_\sigma+$, the equation of gauge-variation \eqref{11} becomes
\begin{equation}
\delta\left(\Omega_{\mathbf{n}\to\mathbf{n}''}(\mathbf{r})-\frac{1}{g}\Lambda(\mathbf{r})\right)=-4\pi,
\end{equation}
the representation of U(1) gauge transformation only permitted as
\begin{equation}\label{15}
\begin{aligned}
\lim_{\mathbf{r}_0\to\mathbf{r}_\sigma-}\psi'(\mathbf{r}_0)&=\lim_{\mathbf{r}_0\to\mathbf{r}_\sigma-}e^{ieg\Omega_{\mathbf{n}\to\mathbf{n}''}(\mathbf{r}_0)}\psi(\mathbf{r}_0),\\[2ex]
\lim_{\mathbf{r}_1\to\mathbf{r}_\sigma+}\psi'(\mathbf{r}_1)&=\lim_{\mathbf{r}_1\to\mathbf{r}_\sigma-}e^{ieg(\Omega_{\mathbf{n}\to\mathbf{n}''}(\mathbf{r}_1)+4\pi)}\psi(\mathbf{r}_0)
\end{aligned}
\end{equation}
where we put the each notation $\mathbf{r}\to\mathbf{r}_\sigma\pm$ as $\mathbf{r}$ approaching to $\mathbf{r}_\sigma$ from the below or above side of area $\sigma$. Now that the representation of gauge $\Lambda(\mathbf{r})$ is single-valued only for $\mathbf{r}\in\mathbf{r}_\sigma$, but  in generally it has $e^{ieg\Omega_{\mathbf{n}\to\mathbf{n}''}(\mathbf{r}_0)}\neq e^{ieg(\Omega_{\mathbf{n}\to\mathbf{n}''}(\mathbf{r}_1)+4\pi)}$. Limits on \eqref{15} only yields trivial identity:
\begin{equation}
\lim_{\mathbf{r}_0\to\mathbf{r}_\sigma-}{\Omega_{\mathbf{n}\to\mathbf{n}''}(\mathbf{r}_0)}=\lim_{\mathbf{r}_1\to\mathbf{r}_\sigma+}{\Omega_{\mathbf{n}\to\mathbf{n}''}(\mathbf{r}_1)+4\pi}.
\end{equation}
So the discussion to Dirac quantization on \cite{A} must be modified.
\section{the half-plane gauge representation}\label{sec:2}
Let us construct the representation of half-plane gauge transformation what makes string rotation:
\begin{equation}\label{17}
\mathbf{n}(0,0,1)\to\mathbf{n}'(0,0,-1),
\end{equation}
\begin{equation}
\sigma=\lim_{r_p', r_q'\to\infty}\boldsymbol{\Gamma}[r_p', r_q']
\end{equation}
with string transforming area $\sigma$ as $xz$-half plane(FIG. \ref{3a}):\noindent
\begin{figure}[!h]
\centering
\begin{subfigure}{.4\textwidth}
\includegraphics[width=1.2\linewidth]{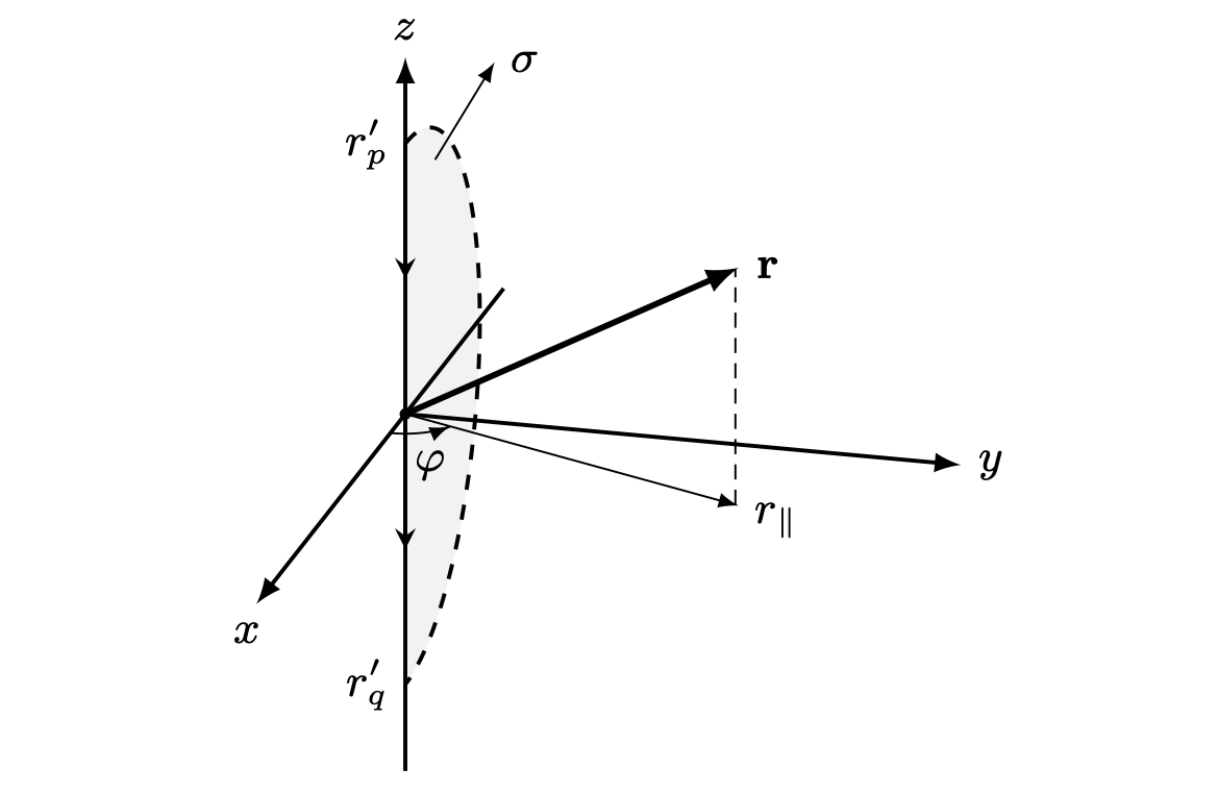}
\caption{}
\end{subfigure}\\
\begin{subfigure}{.4\textwidth}
\includegraphics[width=1.2\linewidth]{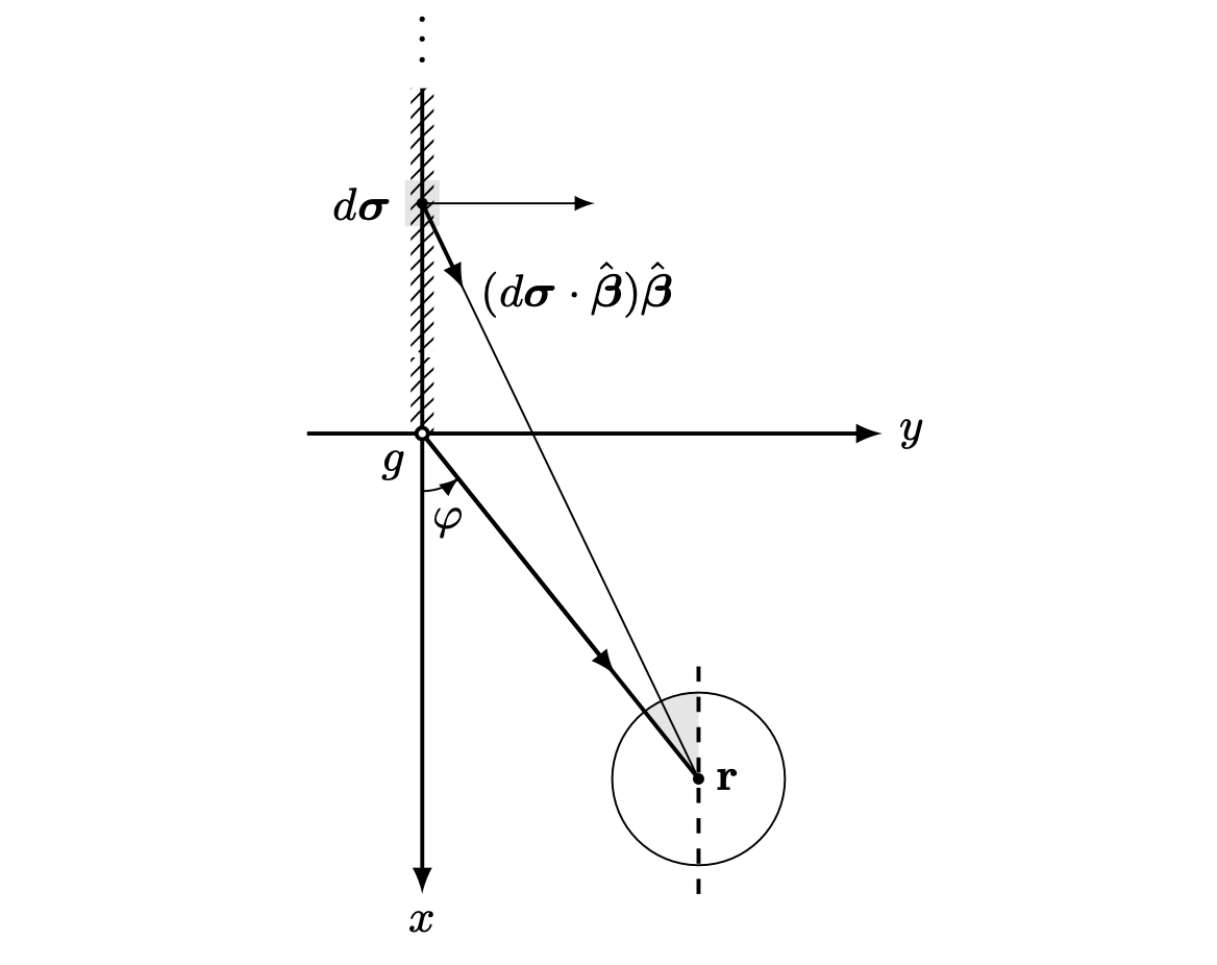}
\caption{}
\end{subfigure}
\caption{String transforming area $\sigma$ on string rotation \eqref{17}.}
\label{3a}
\end{figure}\\
As here we put the variable conversion as $\boldsymbol{\beta}\equiv \mathbf{r}-\mathbf{r}'$ on \eqref{4}, the solid angle $\Omega_{\mathbf{n}\to\mathbf{n}'}(\mathbf{r})$ on this half-plane gauge transformation comes to
 \begin{equation}\label{19}
\Omega_{\mathbf{n}\to\mathbf{n}'}(\mathbf{r})=\int_\sigma \frac{\hat{\boldsymbol{\beta}}\cdot d\boldsymbol{\sigma}}{\beta^2}=\int_{-\pi}^{\varphi-\pi}\int_{0}^{\pi} \sin{\theta'}\ d\theta'd\varphi'=2\varphi.
 \end{equation}
From the representation of gauge \eqref{9}, space of this system is branch-cut by string transforming area $\sigma$, the azimuthal angle has restriction of $\varphi\in (-\pi, \pi)$. If here we put gauge fixing on this string rotation $\mathbf{n}\to\mathbf{n}'$ as
\begin{equation}\label{20}
\Lambda(\mathbf{r})=g\Omega_{\mathbf{n}\to\mathbf{n}'}(\mathbf{r}),
\end{equation}
the gauge transformation of each field on this string rotation comes to
\begin{equation}\label{21}
\begin{aligned}
\psi_{\mathbf{n}}(\mathbf{r})\ \to\ \psi_{\mathbf{n}'}(\mathbf{r})&=e^{ieg\Omega_{\mathbf{n}\to\mathbf{n}'}(\mathbf{r})}\psi_{\mathbf{n}}(\mathbf{r})=e^{2ieg\varphi}\psi_{\mathbf{n}}(\mathbf{r}),\\[2ex]
\mathbf{A}_{\mathbf{n}}(\mathbf{r})\ \to \ \mathbf{A}_{\mathbf{n}'}(\mathbf{r})&=\mathbf{A}_{\mathbf{n}}(\mathbf{r})+\boldsymbol{\nabla}\Lambda(\mathbf{r})=\mathbf{A}_{\mathbf{n}}(\mathbf{r})+\boldsymbol{\nabla}\varphi.
\end{aligned}
\end{equation} 
We shall here confirm that the gauge $\Lambda(\mathbf{r})=2g\varphi$ from calculation of solid angle \eqref{19} exactly gives the half-plane string rotation:
\begin{equation}
\mathbf{A}_{\mathbf{n}}(\mathbf{r})+\boldsymbol{\nabla}\varphi=g\frac{1-\cos{\theta}}{r\sin{\theta}}\ \mathbf{e}_\varphi=\mathbf{A}_{\mathbf{n}'}(\mathbf{r}).
\end{equation}
Let us consider the representation of gauge transformation by generalized string transforming area. By using same method, we also get the expansion of gauge transformation \eqref{21}  for $\varphi_0$-rotated string transforming area $\tilde{\sigma}$ shown on FIG. \ref{4a}:
\begin{figure}[!h]
\centering

\begin{subfigure}{.4\textwidth}
\includegraphics[width=1.2\linewidth]{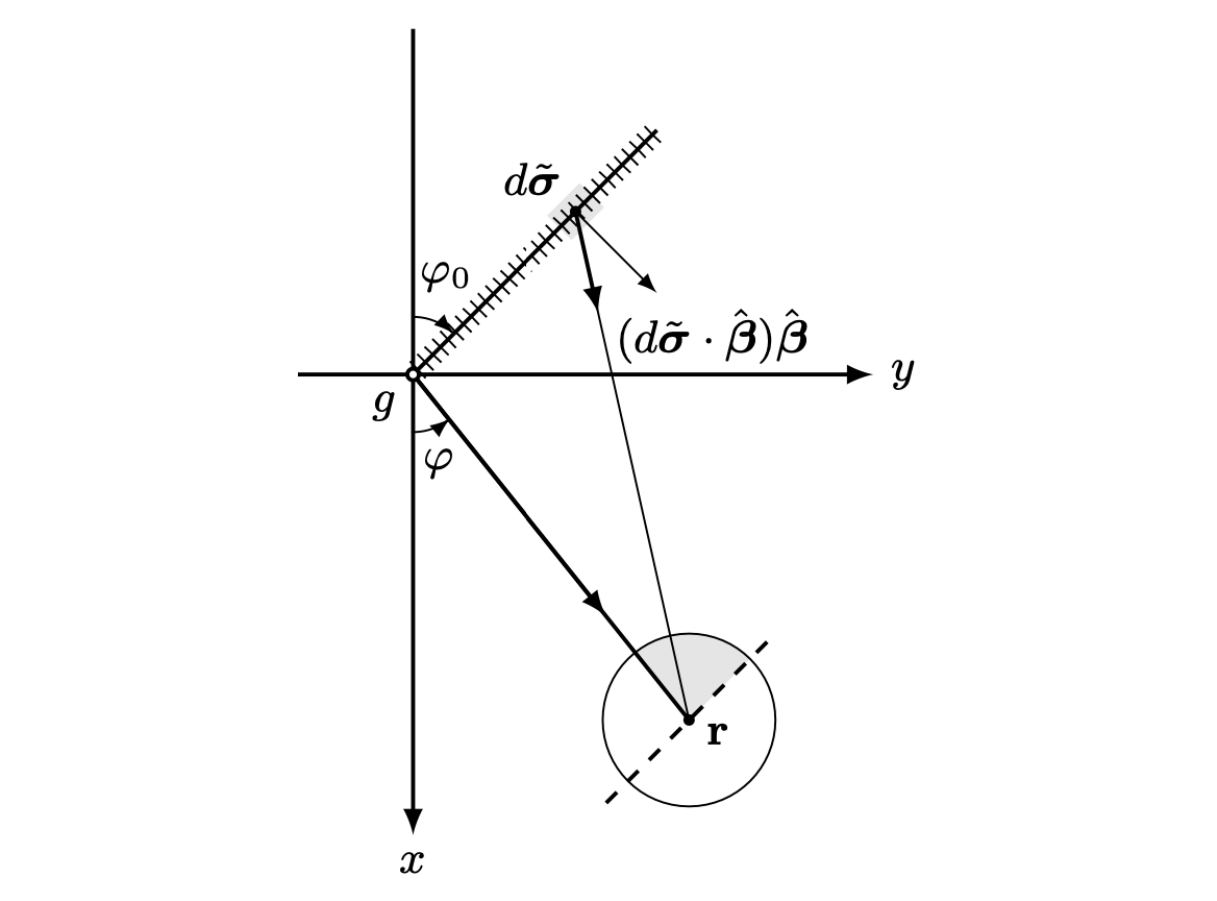}
\caption{}
\label{5.a}
\end{subfigure}\ \ \ \ 
\begin{subfigure}{.4\textwidth}
\includegraphics[width=1.2\linewidth]{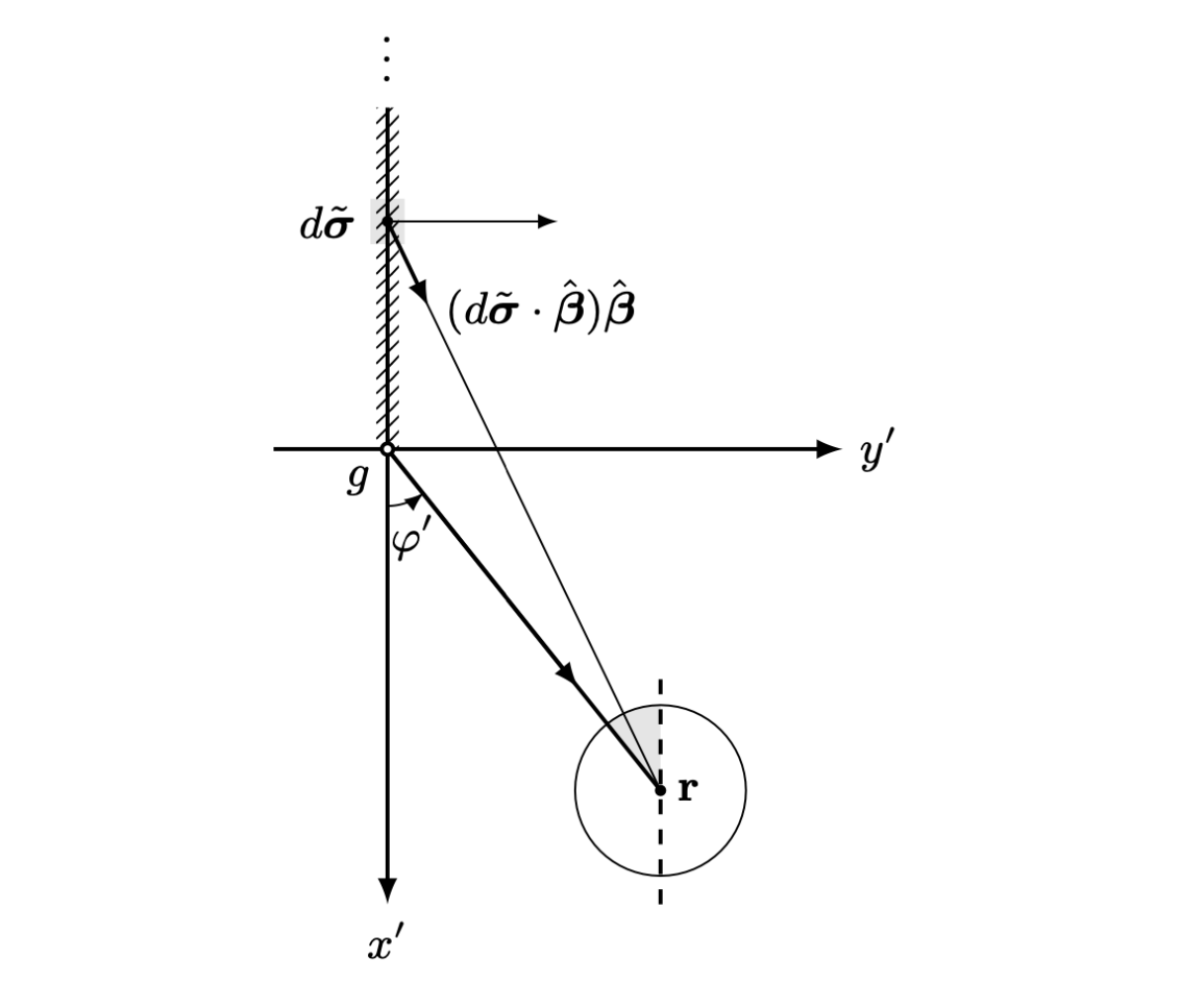}
\caption{}
\label{5.b}
\end{subfigure}
\caption{The rotated string transforming area $\tilde{\sigma}$.}
\label{4a}
\end{figure}\\
This gauge transformation by area $\tilde{\sigma}$ is geometrically equivalent to FIG. \ref{3a}. Hence, here we shall take the proper rotation of coordinates $(x,y)\mapsto(x', y')$ what makes FIG. \ref{5.a} comes to FIG. \ref{5.b}. The azimuthal angle of vector $\mathbf{r}$ on new coordinates $(x', y')$ now comes to
\begin{equation}
\varphi'=\varphi+\varphi_0.
\end{equation}
On the gauge fixing \eqref{20}, it leads the representation of this gauge transformation $\mathbf{n}\to\mathbf{n}'$ by rotated string transforming area $\tilde{\sigma}$ as:
\begin{equation}\label{24}
\begin{aligned}
\psi_{\mathbf{n}}(\mathbf{r})\ \to\ \psi_{\mathbf{n}'}(\mathbf{r})&=e^{2ieg\varphi'}\psi_{\mathbf{n}}(\mathbf{r})=e^{2ieg(\varphi+\varphi_0)}\psi_{\mathbf{n}}(\mathbf{r}),\\[2ex]
\mathbf{A}_{\mathbf{n}}(\mathbf{r})\ \to \ \mathbf{A}_{\mathbf{n}'}(\mathbf{r})&=\mathbf{A}_{\mathbf{n}}(\mathbf{r})+\boldsymbol{\nabla}(\varphi+\varphi_0)=\mathbf{A}_{\mathbf{n}}(\mathbf{r})+\boldsymbol{\nabla}\varphi
\end{aligned}
\end{equation} 
where it takes the gauge $\Lambda(\mathbf{r})=2g(\varphi+\varphi_0)$. This representation is supported by results of \eqref{21}. Also, the gauge fixing is still held after the rotation of coordinates $(x,y)\mapsto(x', y')$. This expression shows that the constant term of phase is related to the direction of string transforming area.
\section{the phase gauge transformation}
Let us form the gauge representation of {\it phase gauge trnasformation} what makes following 2-step string rotation:
\begin{equation}\label{3.3}
u: \ \ \ \mathbf{n}(0,0,0) \ \xrightarrow[U_+(0)]{}\ \mathbf{n}''(0,0,-1)\ \xrightarrow[U_-(\varphi_0)]{}\ \mathbf{n}(0,0,1),
\end{equation}
\begin{equation}
\sigma_+=\lim_{\xi_p', \xi_q'\to\infty}\boldsymbol{\Gamma}_a[\xi_p', \xi_q']\ , \ \ \ \sigma_-=\lim_{\xi_p', \xi_q'\to\infty}\boldsymbol{\Gamma}_b[\xi_q', \xi_p']
\end{equation}
with string transforming area $\sigma_\pm$ as $xz$-half plane(FIG. \ref{6a}):
\begin{widetext}

\begin{figure}[!h ]
\centering
\begin{subfigure}{.4\textwidth}
\begin{tikzpicture}[scale=.7]
\tdplotsetmaincoords{70}{105}
\begin{scope}[tdplot_main_coords]
\fill[gray!10, domain=pi/2:3*pi/2,samples=100,variable=\x]  plot({2.5*cos(\x r)}, {0}, {2.5*sin(\x r)})--(0,0,-2.5)--(0,0,2.5);

 \draw[thick, -latex] (-3.3,0,0)--(5,0,0) node[left, yshift=-2mm, xshift=1.5mm]{\small$x$};
 \draw[thick, -latex] (0,0,0)--(0,4.5,0) node[right]{\small$y$};
 \draw[thick, -latex] (0,0,-3.3)--(0,0,3.3) node[above]{\small$z$};
  \draw[thick,domain=pi/2:3*pi/2,samples=100,variable=\x,dashed]
            plot({2.5*cos(\x r)}, {0}, {2.5*sin(\x r)});
            \draw[-latex, domain=0:1.15,samples=100,variable=\x]
            plot({.5*cos(\x r)}, {.5*sin(\x r)}, {0}) node[right, yshift=-2.7mm, xshift=-3.6mm]{\small$\varphi$};

            \draw[-stealth, thick] (0,0,2.5)--(0,0,1.25);
             \draw[-stealth, thick] (0,0,0)--(0,0,-1.25);
            \fill[black] (0,0,0) circle(.5mm);
            \node at (0,0,2.5) [left]{$\xi_p'$};
               \node at (0,0,-2.5) [left]{$\xi_q'$};
            \draw[-latex] (-1, 0 ,2)--(-3, 0,2.2) node[right]{$\sigma_+$};
            \draw[very thick, -latex] (0,0,0)--(1.5, 3.4, 2.2) node[right]{$\mathbf{r}$};
            \draw[-latex] (0,0,0)--(1.5,3.4,0)node[right, yshift=-1mm]{$r_{\parallel}$};
            \draw[densely dashed](1.5*.99,3.4*.99,0)--(1.5*.99,3.4*.99,2.2*.99);
             \end{scope}
\end{tikzpicture}
\caption{First step gauge transformation $U_+(0)$.}
\label{6.a}
\end{subfigure}\ \ \ \ \ \ \ \ \ \ \ \  \ \ \ \ \ \   \begin{subfigure}{.4\textwidth}
\begin{tikzpicture}[scale=.7]
\tdplotsetmaincoords{70}{105}
\begin{scope}[tdplot_main_coords]
\fill[gray!10, domain=pi/2:3*pi/2,samples=100,variable=\x]  plot({2.5*cos(\x r)*cos(45)}, {2.5*cos(\x r)*sin(45)}, {2.5*sin(\x r)})--(0,0,-2.5)--(0,0,2.5);

 \draw[thick, -latex] (-3.3,0,0)--(5,0,0) node[left, yshift=-2mm, xshift=1.5mm]{\small$x$};
 \draw[thick, -latex] (0,0,0)--(0,4.7,0) node[right]{\small$y$};
 \draw[thick, -latex] (0,0,-3.3)--(0,0,3.3) node[above]{\small$z$};
  \draw[thick,domain=pi/2:3*pi/2,samples=100,variable=\x,dashed]
            plot({2.5*cos(\x r)*cos(45)}, {2.5*cos(\x r)*sin(45)}, {2.5*sin(\x r)});
            \draw[-latex, domain=0:1.15,samples=100,variable=\x]
            plot({.5*cos(\x r)}, {.5*sin(\x r)}, {0}) node[right, yshift=-2.7mm, xshift=-3.6mm]{\small$\varphi$};
             \draw[-latex, domain=pi:-3*pi/4,samples=100,variable=\x]
            plot({.35*cos(\x r)}, {.35*sin(\x r)}, {0}) node[left, yshift=-2.4mm, xshift=-.4mm]{\small$\varphi_0$};

            \draw[-stealth, thick] (0,0,0)--(0,0,1.25);
             \draw[-stealth, thick] (0,0,-2.5)--(0,0,-1.25);
            \fill[black] (0,0,0) circle(.5mm);
            \node at (0,0,2.5) [right]{$\xi_p'$};
               \node at (0,0,-2.5) [right]{$\xi_q'$};
            \draw[-latex] (0, -.2 ,2.3)--(0, -.6,3.2) node[left]{$\sigma_-$};
            \draw[very thick, -latex] (0,0,0)--(1.5, 3.4, 2.2) node[right]{$\mathbf{r}$};
            \draw[-latex] (0,0,0)--(1.5,3.4,0)node[right, yshift=-1mm]{$r_{\parallel}$};
            \draw (0,0,0)--(-2.5/2^.5, -2.5/2^.5,0);
            \fill[black] (-2.5/2^.5, -2.5/2^.5,0) circle(.5mm);
            \draw[densely dashed](1.5*.99,3.4*.99,0)--(1.5*.99,3.4*.99,2.2*.99);
             \end{scope}
\end{tikzpicture}
\caption{Second step gauge transformation $U_-(\varphi_0)$.}
\label{6.b}
\end{subfigure}
\caption{The gauge transforming area $\sigma_\pm$ on {\it phase gauge}.}
\label{6a}
\end{figure}
\begin{figure}[!h]
\centering
\begin{tikzpicture}[scale=.8]

\fill[gray!20] (.15,1.8)--(.15, 2.2)--(-.15,2.2)--(-.15,1.8)--cycle;
\fill[gray!20, rotate=60] (.15,1.8)--(.15, 2.2)--(-.15,2.2)--(-.15,1.8)--cycle;
\fill[pattern=north east lines] (.1,0)--(.1, 3)--(-.1,3)--(-.1,0)--cycle;
\fill[rotate=60, pattern=north east lines] (.1,0)--(.1, 3)--(-.1,3)--(-.1,0)--cycle;
\fill[black](-2*3^.5/2, 1) circle(.5mm);
\draw[thick, rotate=60] (0,0)--(0, 3);

\draw[thick, -latex] (0, 3)--(0, -4) node[below]{\small$x$};
\node at (0,3) [above, yshift=.7mm]{$\vdots$};
\draw[thick, -latex] (0,0)--(2.4*.7, -3*.7);
\draw[thick] (0,0)--(2.4, -3) node[right]{$\mathbf{r}$};
\fill[black] (2.4,-3) circle(.5mm);

\draw[-latex, rotate=60] (0,2)--(-2.3, 2) node[pos=0, above, xshift=5mm]{$d\boldsymbol{\sigma}_-$};
\draw[thick, -latex] (-2*3^.5/2, 1)--(-2*3^.5/2+2*3^.5/2*.2+2.4*.2,1-4*.2) node[yshift=2.2mm,xshift=-5.7mm,left, fill=white]{\small$(d\boldsymbol{\sigma_-}\cdot\hat{\boldsymbol{\beta}})\hat{\boldsymbol{\beta}}$};

\draw[thick, -latex] (-3,0)--(4,0) node[right]{\small$y$};
\draw[-latex] (0,2)--(1.5, 2) node[pos=0, left, xshift=-1mm]{$d\boldsymbol{\sigma}_+$};
\draw[thick, -latex] (0,2)--(0+2.4*.15,2-5*.15) node[right]{$(d\boldsymbol{\sigma}_+\cdot \hat{\boldsymbol{\beta}})\hat{\boldsymbol{\beta}}$};
\fill[white] (0,0) circle(.5mm);
\draw[thick] (0,0) circle(.5mm) ;
\fill[black] (0,2) circle(.5mm);
\draw[-latex, variable=\x, domain=0:.7] plot({0.5*sin(\x r)}, {-.5*cos(\x r)}) node[below, xshift=-.8mm, yshift=-1mm]{\small$\varphi$};
\draw[-latex, variable=\x, domain=pi:-2*pi/3] plot({0.4*sin(\x r)}, {-.4*cos(\x r)}) node[below, xshift=-1.4mm, yshift=-3.5mm]{\small$\varphi_0$};
\end{tikzpicture}
\caption{The Diagramme of string transforming area $\sigma_\pm$ and vector $\boldsymbol{\beta}$ on $xy$ plane. System is branch-cut by area $\sigma_\pm$.}
\label{6aa}
\end{figure} 

\end{widetext}
Since this {\it phase gauge transformation} only holds the direction of Dirac string, it leads the U(1) gauge representation on this string rotation by
\begin{equation}\label{27}
\begin{aligned}
\psi(\mathbf{r})\ &\xrightarrow[u]{}\ \psi'(\mathbf{r})=e^{ie\Lambda_0}\psi(\mathbf{r}),\ \ \ \ \ \ \ \mathbf{A}(\mathbf{r})\ &\xrightarrow[u]{}\ \mathbf{A}'(\mathbf{r})=\mathbf{A}(\mathbf{r})
\end{aligned}
\end{equation}
for a constant phase $\Lambda_0$. With the half-plane gauge representation derived on \ref{sec:2} now we come to the representation of phase $\Lambda_0$ on U(1) gauge \eqref{27}. By taking the representation of {\it half-plane gauge representation} for each $U_+(0)=e^{2ieg\varphi}$, $U_-(\varphi_0)=e^{-2ieg(\varphi+\varphi_0)}$ driven on \eqref{21} and \eqref{24} on gauge fixing \eqref{20}, we thus get the representation of phase gauge transformation $u$ as:
\begin{equation}\label{28}
u: \ \ \psi(\mathbf{r})\ \xrightarrow[u]{}\ U_+(0)U_-(\varphi_0)\psi(\mathbf{r})=e^{-2ieg\varphi_0}\psi(\mathbf{r}).
\end{equation}
We may also take different approach to this {\it phase gauge transformation}.  If we remain the term of boundary path $G_{\xi_q'\to\xi_p'}^a$, $G_{\xi_p'\to\xi_q'}^b$ (each dashed path shown on FIG. \ref{6a}), the integral representation of gauge transformation  \eqref{7} comes to
\begin{widetext}
\begin{equation}\label{29}
\begin{aligned}
\mathbf{0}&=\lim_{\xi_p', \xi_q'\to\infty}\left(g\sum_{i}\int_{\partial\boldsymbol{\Gamma}_i}\frac{(\mathbf{r}-\mathbf{r}')\times d\mathbf{r}'}{|\mathbf{r}-\mathbf{r}'|^3}-g\int_{G_{\xi_q'\to\xi_p'}^a+G_{\xi_p'\to\xi_q'}^b}\frac{(\mathbf{r}-\mathbf{r}')\times d\mathbf{r}'}{|\mathbf{r}-\mathbf{r}'|^3}\right)\\[2ex]
&=g\boldsymbol{\nabla}\left(\Omega_{\mathbf{n}\to\mathbf{n}'}(\mathbf{r})+\Omega_{\mathbf{n}'\to\mathbf{n}}(\mathbf{r})-\lim_{\xi_p', \xi_q'\to\infty}\int_{\mathcal{G}_{\xi'}}\frac{(\mathbf{r}-\mathbf{r}')\cdot d\boldsymbol{\sigma}'}{|\mathbf{r}-\mathbf{r}'|^3}\right)
\end{aligned}
\end{equation}
\end{widetext}
with $i=a, b$ on $\mathbf{r}\not\in \sigma_\pm.$ So the relation \eqref{29} also leads
\begin{equation}\label{30}
\Omega_{\mathbf{n}\to\mathbf{n}'}(\mathbf{r})+\Omega_{\mathbf{n}'\to\mathbf{n}}(\mathbf{r})-\lim_{\xi_p', \xi_q'\to\infty}\int_{\mathcal{G}_{\xi'}}\frac{(\mathbf{r}-\mathbf{r}')\cdot d\boldsymbol{\sigma}'}{|\mathbf{r}-\mathbf{r}'|^3}=\alpha_0
\end{equation}
for constant $\alpha_0$. Here we denote the area $\mathcal{G}_{\xi'}$ as a surface\footnote{The boundary term on \eqref{29} exactly vanished:
\begin{equation*}
g\int_{G_{\xi_q'\to\xi_p'}^a+G_{\xi_p'\to\xi_q'}^b}\frac{(\mathbf{r}-\mathbf{r}')\times d\mathbf{r}'}{|\mathbf{r}-\mathbf{r}'|^3}\ \xrightarrow[\xi_p', \xi_q'\to\infty]{}\ 0.
\end{equation*}} with boundary $\partial \mathcal{G}_{\xi'}=G_{\xi_q'\to\xi_p'}^a+G_{\xi_p'\to\xi_q'}^b$  shown on FIG. \ref{7a}:
\begin{figure}[!h]
\centering
\begin{tikzpicture}[scale=.8]
\tdplotsetmaincoords{48}{105}
\begin{scope}[tdplot_main_coords, sphere segment/.style args={%
phi from #1 to #2 and theta from #3 to #4 and radius #5}{insert path={%
 plot[variable=\x,smooth,domain=#2:#1] 
 (xyz spherical cs:radius=#5,longitude=\x,latitude=#3)
 -- plot[variable=\x,smooth,domain=#3:#4] 
 (xyz spherical cs:radius=#5,longitude=#1,latitude=\x)
 --plot[variable=\x,smooth,domain=#1:#2] 
 (xyz spherical cs:radius=#5,longitude=\x,latitude=#4)
 -- plot[variable=\x,smooth,domain=#4:#3] 
 (xyz spherical cs:radius=#5,longitude=#2,latitude=\x)}},
>=stealth,declare function={f(\x)=exp(-2+0.5*\x);}]
\fill[gray!27, domain=pi/2:3*pi/2,samples=100,variable=\x]  plot({2.5*cos(\x r)*cos(45)}, {2.5*cos(\x r)*sin(45)}, {2.5*sin(\x r)})--(0,0,-2.5)--(0,0,2.5);
\fill[gray!15, domain=pi/2:3*pi/2,samples=100,variable=\x]  plot({2.5*cos(\x r)}, {0}, {2.5*sin(\x r)})--(0,0,-2.5)--(0,0,2.5);
\fill[black] (0,0,2.5) circle(.5mm);
\fill[black] (0,0,-2.5) circle(.5mm);
  \shadedraw[tdplot_screen_coords,ball color = white, opacity=0.28] (0,0) circle (\rvec);
                     \draw[color=white,semithick, fill=white,fill opacity=1,
 sphere segment={phi from -90 to -90-45 and theta from 46.5 to 90 and radius 2.5}] ; 

 \draw[thick, -latex] (-3.3,0,0)--(5,0,0) node[left, yshift=-2mm, xshift=1.5mm]{\small$x$};
 \draw[thick, -latex] (0,0,0)--(0,5,0) node[right]{\small$y$};
 \draw[thick, -latex] (0,0,-3)--(0,0,4.1) node[above]{\small$z$};
  \draw[very thick,domain=pi/2:3*pi/2,samples=100,variable=\x, dashed]
            plot({2.5*cos(\x r)*cos(45)}, {2.5*cos(\x r)*sin(45)}, {2.5*sin(\x r)});
             \draw[-stealth, very thick,domain=pi/2:2.3*pi/2,samples=100,variable=\x, dashed]
            plot({2.5*cos(\x r)*cos(45)}, {2.5*cos(\x r)*sin(45)}, {2.5*sin(\x r)});
            \draw[-latex, domain=0:1.15,samples=100,variable=\x]
            plot({.5*cos(\x r)}, {.5*sin(\x r)}, {0}) node[right, yshift=-2.7mm, xshift=-3.6mm]{\small$\varphi$};
             \draw[-latex, domain=pi:-3*pi/4,samples=100,variable=\x]
            plot({.35*cos(\x r)}, {.35*sin(\x r)}, {0}) node[left, yshift=-2.4mm, xshift=-.4mm]{\small$\varphi_0$};
 \draw[thick,domain=pi/2:1.9*pi/2,samples=100,variable=\x, dashed]
            plot({2.5*cos(\x r)}, {0}, {2.5*sin(\x r)});
             \draw[stealth-, thick,domain=1.9*pi/2:3*pi/2,samples=100,variable=\x, dashed]
            plot({2.5*cos(\x r)}, {0}, {2.5*sin(\x r)});
  \draw[-latex] (-1, 0 ,1.8)--(-3, 0,2) node[right]{\small$\boldsymbol{\Gamma}_a$};
            \draw[-stealth, very thick] (0,0,0)--(0,0,1.25);
             \draw[-stealth, very thick] (0,0,-2.5)--(0,0,-1.25);
              \draw[very thick] (0,0,-2.5)--(0,0,2.5);
            \fill[black] (0,0,0) circle(.5mm);
            \node at (0,0,2.5) [yshift=7mm, left]{$\xi_p'$};
               \node at (0,0,-2.5) [left, yshift=-2mm]{$\xi_q'$};
            \draw[-latex] (0, -.2 -.8,2.3-.3)--(0, -.6-1,3.7-.7) node[left]{\small$\boldsymbol{\Gamma}_b$};
            \draw[semithick, -latex] (0,0,0)--(1.5, 3.4, 2.2) node[right]{$\mathbf{r}$};
            \draw[-latex] (0,0,0)--(1.5,3.4,0)node[right, yshift=-1mm]{$r_{\parallel}$};
            \draw (0,0,0)--(-2.5/2^.5, -2.5/2^.5,0);
                     \draw[densely dashed](1.5*.99,3.4*.99,0)--(1.5*.99,3.4*.99,2.2*.99);
                   \draw[thick, color=gray!60, variable=\x, domain=pi:-.75*pi, samples=100] plot({2.5*cos(\x r)}, {2.5*sin(\x r)}, {0});
                     \draw[thick, color=gray!60, variable=\x, domain=pi:.25*pi, samples=100, -latex] plot({2.5*cos(\x r)}, {2.5*sin(\x r)}, {0});
                     \draw[-latex] (2.3, -.8, -.7)--(2.9, -1.3, -.7) node[left]{\small$\mathcal{G}_{\xi'}$};
             \end{scope}
\end{tikzpicture}
\caption{Path on integral \eqref{29} and \eqref{30}.}
\label{7a}
\end{figure}\\
Now we shall evaluate the constant $\alpha_0$ using Gauss' law on integral \eqref{30} by
 \begin{equation}
 \begin{aligned}
 \alpha_0&=\lim_{\xi_p', \xi_q'\to\infty}\oint_{\boldsymbol{\Gamma}_a[\xi_p', \xi_q']+\boldsymbol{\Gamma}_b[\xi_q', \xi_p']-\mathcal{G}_\xi'} \frac{(\mathbf{r}-\mathbf{r}')\cdot d\boldsymbol{\sigma}'}{|\mathbf{r}-\mathbf{r}'|^3}\\[2ex]
 &=\lim_{\xi_p', \xi_q'\to\infty}-\int_{\text{internal space}} \nabla'^2\left(\frac{1}{|\mathbf{r}-\mathbf{r}'|}\right)\ d\Sigma'\\[2ex]
 &=4\pi\theta_{\pi}(\varphi_0-\varphi)
 \end{aligned}
 \end{equation}
with azimuthal step function $\theta_\pi(\varphi_0-\varphi)$:
\begin{equation}\label{31}
\theta_\pi(\varphi_0-\varphi)=\begin{cases}
1 \ \ \ \ \ \ \ &(\varphi_0<\pi-\varphi)\\[2ex]
0 \ \ \ \ \ \ \ &(\varphi_0>\pi-\varphi)
\end{cases}.                                                                                                                              
\end{equation}
Combining the \eqref{31} and \eqref{30}, we may also get the integral form of solid angle $\Omega_{\mathbf{n}\to\mathbf{n}'}(\mathbf{r})+\Omega_{\mathbf{n}'\to\mathbf{n}}(\mathbf{r})$. Since the Dirac monopole is abelian, it leads the integral form of phase gauge transformation $u$ as
\begin{widetext}
\begin{equation}\label{33}
u: \ \ \ U_+(0)U_-(\varphi_0)=\exp\left(4\pi ieg\theta_\pi(\varphi_0-\varphi)+\lim_{\xi_p', \xi_q'\to\infty}ieg\int_{\mathcal{G}_{\xi'}}\frac{(\mathbf{r}-\mathbf{r}')\cdot d\boldsymbol{\sigma}'}{|\mathbf{r}-\mathbf{r}'|^3}\right)
\end{equation}
on gauge fixing \eqref{20}. Comparing these two representation \eqref{28} and \eqref{33} of phase gauge transformation $u$ , we get
\begin{equation}\label{34}
e^{-2ieg\varphi_0}=\exp\left(4\pi ieg\theta_\pi(\varphi_0-\varphi)+\lim_{\xi_p', \xi_q'\to\infty}ieg\int_{\mathcal{G}_{\xi'}}\frac{(\mathbf{r}-\mathbf{r}')\cdot d\boldsymbol{\sigma}'}{|\mathbf{r}-\mathbf{r}'|^3}\right).
\end{equation}
\end{widetext}
\section{The Dirac Quantization}
In this section, we investigate geometrical approach to Dirac quantization under phase gauge transformation. The representation\footnote{Here we consider the phase gauge transformation with path on FIG. \ref{6a}.} of { phase gauge representation} $u$ is given by 
\begin{equation}\label{35}
\begin{aligned}
\psi(\mathbf{r})\ &\xrightarrow[u]{}\ \psi'(\mathbf{r})=e^{-2ieg\varphi_0}\psi(\mathbf{r}),\\[2ex]
\mathbf{A}(\mathbf{r})\ &\xrightarrow[u]{}\ \mathbf{A}'(\mathbf{r})=\mathbf{A}(\mathbf{r})
\end{aligned}
\end{equation}
on gauge fixing \eqref{20}. So the phase gauge transformation only gives the constant phase difference to $\psi(\mathbf{r})$. The key idea to Dirac quantization consists in finding a proper path for phase gauge transformation $\slashed{u}$ what transforms
\begin{equation}\label{36}
\begin{aligned}
\psi(\mathbf{r})\ &\xrightarrow[\slashed{u}]{}\ \psi(\mathbf{r}), \ \ \mathbf{A}(\mathbf{r})\ &\xrightarrow[\slashed{u}]{}\ \mathbf{A}(\mathbf{r})
\end{aligned}
\end{equation}
Now we shall consider following two paths shown on FIG. \ref{8a} and FIG. \ref{9a}:
\begin{figure}[!h]
\centering
\begin{subfigure}{0.4\textwidth}
\begin{tikzpicture}[scale=.8]
\tdplotsetmaincoords{48}{105}
\begin{scope}[tdplot_main_coords, sphere segment/.style args={%
phi from #1 to #2 and theta from #3 to #4 and radius #5}{insert path={%
 plot[variable=\x,smooth,domain=#2:#1] 
 (xyz spherical cs:radius=#5,longitude=\x,latitude=#3)
 -- plot[variable=\x,smooth,domain=#3:#4] 
 (xyz spherical cs:radius=#5,longitude=#1,latitude=\x)
 --plot[variable=\x,smooth,domain=#1:#2] 
 (xyz spherical cs:radius=#5,longitude=\x,latitude=#4)
 -- plot[variable=\x,smooth,domain=#4:#3] 
 (xyz spherical cs:radius=#5,longitude=#2,latitude=\x)}},
>=stealth,declare function={f(\x)=exp(-2+0.5*\x);}]

\fill[gray!27, domain=pi/2:3*pi/2,samples=100,variable=\x]  plot({2.5*cos(\x r)*cos(45)}, {2.5*cos(\x r)*sin(45)}, {2.5*sin(\x r)})--(0,0,-2.5)--(0,0,2.5);
\fill[gray!15, domain=pi/2:3*pi/2,samples=100,variable=\x]  plot({2.5*cos(\x r)}, {0}, {2.5*sin(\x r)})--(0,0,-2.5)--(0,0,2.5);
\fill[black] (0,0,2.5) circle(.5mm);
\fill[black] (0,0,-2.5) circle(.5mm);

 \draw[thick, -latex] (-3.3,0,0)--(5,0,0) node[left, yshift=-2mm, xshift=1.5mm]{\small$x$};
 \draw[thick, -latex] (0,0,0)--(0,4.5,0) node[right]{\small$y$};
 \draw[thick, -latex] (0,0,-3)--(0,0,4.1) node[above]{\small$z$};
  \draw[very thick,domain=pi/2:3*pi/2,samples=100,variable=\x, dashed]
            plot({2.5*cos(\x r)*cos(45)}, {2.5*cos(\x r)*sin(45)}, {2.5*sin(\x r)});
             \draw[-stealth, very thick,domain=pi/2:2.3*pi/2,samples=100,variable=\x, dashed]
            plot({2.5*cos(\x r)*cos(45)}, {2.5*cos(\x r)*sin(45)}, {2.5*sin(\x r)});
            \draw[-latex, domain=0:1.15,samples=100,variable=\x]
            plot({.5*cos(\x r)}, {.5*sin(\x r)}, {0}) node[right, yshift=-2.7mm, xshift=-3.6mm]{\small$\varphi$};
             \draw[-latex, domain=pi:-3*pi/4,samples=100,variable=\x]
            plot({.35*cos(\x r)}, {.35*sin(\x r)}, {0}) node[left, yshift=-2.4mm, xshift=-.4mm]{\small$\varphi_0$};
 \draw[thick,domain=pi/2:1.9*pi/2,samples=100,variable=\x, dashed]
            plot({2.5*cos(\x r)}, {0}, {2.5*sin(\x r)});
             \draw[stealth-, thick,domain=1.9*pi/2:3*pi/2,samples=100,variable=\x, dashed]
            plot({2.5*cos(\x r)}, {0}, {2.5*sin(\x r)});
  \draw[-latex] (-1, 0 ,1.8)--(-3, 0,2) node[right]{\small$\boldsymbol{\Gamma}_a$};
            \draw[-stealth, very thick] (0,0,0)--(0,0,1.25);
             \draw[-stealth, very thick] (0,0,-2.5)--(0,0,-1.25);
              \draw[very thick] (0,0,-2.5)--(0,0,2.5);
            \fill[black] (0,0,0) circle(.5mm);
            \node at (0,0,2.5) [yshift=7mm, left]{$\xi_p'$};
               \node at (0,0,-2.5) [left, yshift=-2mm]{$\xi_q'$};
            \draw[-latex] (0, -.2 -.8,2.3-.3)--(0, -.6-1,3.7-.7) node[left]{\small$\boldsymbol{\Gamma}_b(\varphi_0)$};
            \draw[semithick, -latex] (0,0,0)--(1.5, 3.4, 2.2) node[right]{$\mathbf{r}$};
            \draw[-latex] (0,0,0)--(1.5,3.4,0)node[right, yshift=-1mm]{$r_{\parallel}$};
            \draw (0,0,0)--(-2.5/2^.5, -2.5/2^.5,0);
                     \draw[densely dashed](1.5*.99,3.4*.99,0)--(1.5*.99,3.4*.99,2.2*.99);
                   \draw[thick, color=gray!60, variable=\x, domain=pi:-.75*pi, samples=100] plot({2.5*cos(\x r)}, {2.5*sin(\x r)}, {0});
                     \draw[thick, color=gray!60, variable=\x, domain=pi:.25*pi, samples=100, -latex] plot({2.5*cos(\x r)}, {2.5*sin(\x r)}, {0});
             \end{scope}
\end{tikzpicture}
\end{subfigure}
\caption{Integral path of phase gauge transformation on path $\displaystyle \partial \mathbf{D}_-=\lim_{\varphi_0\to 2\pi-}\partial\boldsymbol{\Gamma}_1+\partial\boldsymbol{\Gamma}_2(\varphi_0)$.}
\label{8a}
\end{figure}\\
We may investigate the representation of phase gauge transformation on path $\partial\mathbf{D}_-$ shown FIG. \ref{8a} using \eqref{33}. Since each azimuthal step function $\theta_{\pi}(\varphi_0-\varphi)$ and boundary integral term on path $\partial\mathbf{D}_-$ goes to
\begin{equation}\label{37}
\theta_\pi(\varphi_0-\varphi)=1, \ \ \lim_{\xi_p', \xi_q'\to\infty}\int_{\mathcal{G}_{\xi'}}\frac{(\mathbf{r}-\mathbf{r}')\cdot d\boldsymbol{\sigma}'}{|\mathbf{r}-\mathbf{r}'|^3}=-4\pi,
\end{equation}
we thus have the representation of phase gauge transformation on path $\partial\mathbf{D}_-$ by
\begin{equation}\label{38}
\begin{aligned}
{u}_{D-}: \ \ \ &\exp\Bigg(4\pi ieg\theta_\pi(\varphi_0-\varphi)\\[2ex]
&\left. \ \ \ \ \ +\lim_{\xi_p', \xi_q'\to\infty}ieg\int_{\mathcal{G}_{\xi'}}\frac{(\mathbf{r}-\mathbf{r}')\cdot d\boldsymbol{\sigma}'}{|\mathbf{r}-\mathbf{r}'|^3}\right)_{D_-}=\mathbf{1}.
\end{aligned}
\end{equation}
Similarly, we also consider the phase gauge transformation on path $\partial \mathbf{D}_+$:
\begin{figure}[!h]
\centering
\begin{subfigure}{0.4\textwidth}
\begin{tikzpicture}[scale=.8]
\tdplotsetmaincoords{48}{105}
\begin{scope}[tdplot_main_coords, sphere segment/.style args={%
phi from #1 to #2 and theta from #3 to #4 and radius #5}{insert path={%
 plot[variable=\x,smooth,domain=#2:#1] 
 (xyz spherical cs:radius=#5,longitude=\x,latitude=#3)
 -- plot[variable=\x,smooth,domain=#3:#4] 
 (xyz spherical cs:radius=#5,longitude=#1,latitude=\x)
 --plot[variable=\x,smooth,domain=#1:#2] 
 (xyz spherical cs:radius=#5,longitude=\x,latitude=#4)
 -- plot[variable=\x,smooth,domain=#4:#3] 
 (xyz spherical cs:radius=#5,longitude=#2,latitude=\x)}},
>=stealth,declare function={f(\x)=exp(-2+0.5*\x);}]

\fill[gray!15, domain=pi/2:3*pi/2,samples=100,variable=\x]  plot({2.5*cos(\x r)}, {0}, {2.5*sin(\x r)})--(0,0,-2.5)--(0,0,2.5);
\fill[black] (0,0,2.5) circle(.5mm);
\fill[black] (0,0,-2.5) circle(.5mm);
 \draw[thick,domain=pi/2:1.9*pi/2,samples=100,variable=\x, dashed]
            plot({2.5*cos(\x r)}, {0}, {2.5*sin(\x r)});
             \draw[stealth-, thick,domain=1.9*pi/2:3*pi/2,samples=100,variable=\x, dashed]
            plot({2.5*cos(\x r)}, {0}, {2.5*sin(\x r)});
   \draw[thick, color=black, variable=\x, domain=pi:pi-.25*pi, samples=100] plot({2.5*cos(\x r)}, {2.5*sin(\x r)}, {0});
                     \draw[thick, color=black, variable=\x, domain=pi:pi-.125*pi, samples=100, -latex] plot({2.5*cos(\x r)}, {2.5*sin(\x r)}, {0});
\fill[gray!37, domain=pi/2:3*pi/2,samples=100,variable=\x, opacity=.6]  plot({2.5*cos(\x r)*cos(-45)}, {2.5*cos(\x r)*sin(-45)}, {2.5*sin(\x r)})--(0,0,-2.5)--(0,0,2.5);
 \draw[thick, -latex] (-3.3,0,0)--(5,0,0) node[left, yshift=-2mm, xshift=1.5mm]{\small$x$};
 \draw[thick, -latex] (0,0,0)--(0,4.5,0) node[right]{\small$y$};
 \draw[thick, -latex] (0,0,-3)--(0,0,4.1) node[above]{\small$z$};
  \draw[very thick,domain=pi/2:3*pi/2,samples=100,variable=\x, dashed]
            plot({2.5*cos(\x r)*cos(-45)}, {2.5*cos(\x r)*sin(-45)}, {2.5*sin(\x r)});
             \draw[-stealth, very thick,domain=pi/2:2.3*pi/2,samples=100,variable=\x, dashed]
            plot({2.5*cos(\x r)*cos(-45)}, {2.5*cos(\x r)*sin(-45)}, {2.5*sin(\x r)});
            \draw[-latex, domain=0:1.15,samples=100,variable=\x]
            plot({.5*cos(\x r)}, {.5*sin(\x r)}, {0}) node[right, yshift=-2.7mm, xshift=-3.6mm]{\small$\varphi$};
             \draw[-latex, domain=pi:pi-pi/4,samples=100,variable=\x]
            plot({.5*cos(\x r)}, {.5*sin(\x r)}, {0}) node[right, yshift=3.5mm, xshift=.7mm]{\small$\varphi_0$};
  \draw[-latex] (-1, 0.1 ,2)--(-3, 0,2) node[right]{\small$\boldsymbol{\Gamma}_a$};
            \draw[-stealth, very thick] (0,0,0)--(0,0,1.25);
             \draw[-stealth, very thick] (0,0,-2.5)--(0,0,-1.25);
              \draw[very thick] (0,0,-2.5)--(0,0,2.5);
            \fill[black] (0,0,0) circle(.5mm);
            \node at (0,0,2.5) [yshift=7mm, left]{$\xi_p'$};
               \node at (0,0,-2.5) [left, yshift=-2mm]{$\xi_q'$};
            \draw[-latex] (-.3, .2 +1,2.3)--(-.3, .6+2,3.7) node[right]{\small$\boldsymbol{\Gamma}_b(\varphi_0)$};
            \draw[semithick, -latex] (0,0,0)--(1.5, 3.4, 2.2) node[right]{$\mathbf{r}$};
            \draw[-latex] (0,0,0)--(1.5,3.4,0)node[right, yshift=-1mm]{$r_{\parallel}$};
            \draw (0,0,0)--(-2.5/2^.5, 2.5/2^.5,0);
                     \draw[densely dashed](1.5*.99,3.4*.99,0)--(1.5*.99,3.4*.99,2.2*.99);

             \end{scope}
\end{tikzpicture}
\caption{$\varphi_0\to 0+$}
\end{subfigure}
\caption{Integral path of phase gauge transformation on path $\displaystyle \partial \mathbf{D}_+=\lim_{\varphi_0\to 0+}\partial\boldsymbol{\Gamma}_1+\partial\boldsymbol{\Gamma}_2(\varphi_0)$.}
\label{9a}
\end{figure}\\
On the phase gauge transformation with path $\partial\mathbf{D}_+$, \eqref{37} comes to
\begin{equation}
\theta_\pi(\varphi_0-\varphi)=0, \ \ \lim_{\xi_p', \xi_q'\to\infty}\int_{\mathcal{G}_{\xi'}}\frac{(\mathbf{r}-\mathbf{r}')\cdot d\boldsymbol{\sigma}'}{|\mathbf{r}-\mathbf{r}'|^3}=0,
\end{equation}
then we also have the representation of phase gauge transformation on path $\partial\mathbf{D}_-$:
\begin{equation}
\begin{aligned}
{u}_{D+}: \ \ \ &\exp\Bigg(4\pi ieg\theta_\pi(\varphi_0-\varphi)\\[2ex]
&\left. \ \ \ \ \ +\lim_{\xi_p', \xi_q'\to\infty}ieg\int_{\mathcal{G}_{\xi'}}\frac{(\mathbf{r}-\mathbf{r}')\cdot d\boldsymbol{\sigma}'}{|\mathbf{r}-\mathbf{r}'|^3}\right)_{D+}=\mathbf{1}.
\end{aligned}
\end{equation}
Thus, here we confirm that both path $\partial\mathbf{D}_-, \partial\mathbf{D}_+$ of phase transformation gives gauge transformation \eqref{36}. So the proper path for phase gauge transformation $\slashed{u}$ what transforms like \eqref{36} is given by $\partial\mathbf{D}_-, \partial \mathbf{D}_+$.  Also, the path $\partial\mathbf{D}_-$ leads to Dirac quantization from the gauge equation \eqref{34} and gauge representation \eqref{38}:
\begin{equation}
u_{D-}: \ \ \ \lim_{\varphi_0\to 2\pi -}e^{-2ieg\varphi_0}=e^{-4\pi ieg}=1,
\end{equation}
which leads
\begin{equation}\label{42}
eg=\frac{l}{2}, \ \ \ \ (l=\pm 1, \pm 2, \cdots).
\end{equation}
Here we notice that Dirac quantization \eqref{42} is derived geometrically under U(1) gauge invariance without a supposing of canonical quantization $[\hat{q}_i, \hat{p}_j]=i\delta_{ij}$.

\section{Discussion}
In the previous section, we have seen that only the U(1) gauge invariance leads to Dirac quantization condition $eg=l/2$ through the geometric tools. This geometrical approach to Dirac quantization \eqref{42} gives some meaningful interpretation about a physical meaning of gauge representation followed. 

The representation of U(1) gauge \eqref{13} now becomes single-valued quantity:
\begin{equation}
U_1=e^{ieg(\Omega_{\mathbf{n}\to\mathbf{n}'}(\mathbf{r})+4\pi)}=e^{ieg\Omega_{\mathbf{n}\to\mathbf{n}'}(\mathbf{r})}=U_2
\end{equation}
from the result of Dirac quantization \eqref{42}. Thus the gauge of Dirac monopole have single-valued representation everywhere. It is only supported by Dirac quantization \eqref{42}. The single-valued gauge representation is not a trivial postulate, only the results derived by Dirac quantization condition $eg=l/2.$

On a half-plane gauge transformation(section \ref{sec:2}), even we shall add the constant gauge $\Lambda(\mathbf{r})\to \Lambda(\mathbf{r})+\Lambda_0$, the gauge transformation of field $\mathbf{A}(\mathbf{r})$ does not changed. But the phase of field $\psi(\mathbf{r})$ is changed, this phase difference is related to the azimuthal angle $\varphi_0$ of string transforming area(see FIG. \ref{4a}):
\begin{equation}\label{44}
\varphi_0=\frac{\Lambda_0}{2g}-\frac{\pi n}{eg} \in (-\pi, \pi)
\end{equation}
for a proper integer $n$. Here we get the relation \eqref{44} by comparing $e^{ie\Lambda(\mathbf{r})}\to e^{ie(\Lambda(\mathbf{r})+\Lambda_0)}$ and the representation of gauge transformation \eqref{24}.

	

\begin{thebibliography}{9}
	\bibitem{Brandt} R. A. Brandt, J. R. Primack, {\it Dirac monopole theory with and without strings}, \href{https://journals.aps.org/prd/abstract/10.1103/PhysRevD.15.1175}{{\it Phys. Rev.} {\bf{D 15}} (1977) 1175}. 
		\bibitem{taro} P. A. M. Dirac, {\it Quantised singularities in the electromagnetic field}, \href{https://royalsocietypublishing.org/doi/10.1098/rspa.1931.0130}{{\it Proceedings of the Royal Society of London. Series A }{\bf 133} (1931) 60-72}.
				\bibitem{Lee} D. G. Boulware, L. S. Brown, R. N. Cahn, S. D. Ellis, C. Lee, {\it Scattering on magnetic charge}, \href{https://journals.aps.org/prd/abstract/10.1103/PhysRevD.14.2708}{{\it Phys. Rev. D} {\bfseries 14} (1976) 2708-2727}.
				\bibitem{monopole} Yakov M. Shnir, {\it Magnetic Monopoles, Springer} (2010).
		\bibitem{DD} M. Blagojevi{\'c}, P. Senjanovi{\'c}, {\it The quantum field theory of electric and magnetic charge}, \href{https://www.sciencedirect.com/science/article/pii/0370157388900981}{{\it Phys. Rep.} {\bf 157} (1988) 233}.
		\bibitem{A} A. Frenkel, P. Harask{\'o}, {\it Invariance properties of the Dirac monopole}, \href{https://www.sciencedirect.com/science/article/pii/0003491677902421}{{\it Ann. Phys. (N. Y.) {\bf 105} (1977) 288}}.
		
	\end{thebibliography}
\end{document}